\documentclass[twocolumn,aps,amsmath,amssymb,prl]{revtex4}
\usepackage{graphicx}
\usepackage{bm}         
\usepackage{color}         
\usepackage{amsmath}
\usepackage{units}
\usepackage{bbm}
\date{\today}
\begin{document}

\title{Non-trivial Chern numbers in three-terminal Josephson junctions}

\author{Julia S. Meyer and Manuel Houzet}
\affiliation{Univ.~Grenoble Alpes, INAC-PHELIQS, F-38000 Grenoble, France}
\affiliation{CEA, INAC-PHELIQS, F-38000 Grenoble, France}

\begin{abstract}
{Recently it has been predicted that the Andreev bound state spectrum of 4-terminal Josephson junctions may possess zero-energy Weyl singularities. Using one superconducting phase as a control parameter, these singularities are associated with topological transitions between time-reversal symmetry broken phases with different Chern numbers. Here we show that such topological transitions may also be tuned with a magnetic flux through the junction area in a 3-terminal geometry.}
\end{abstract}

\maketitle

{\em Introduction.} Over the past few years topological properties of matter have been investigated in a number of different contexts. Starting with gapped topological phases in insulators and superconductors~\cite{review-topo1,review-topo2}, more recently gapless topological phases such as Weyl semimetals have been discovered~\cite{weyl, w-exp1,w-exp2,w-exp3}. While new materials have allowed one to realize a few of such phases, others may require combining different properties in hybrid structures. Another route to enlarging the number of accessible phases is the study of time-periodic systems, which may host so-called Floquet topological phases~\cite{floquet1,floquet2}. 

Alternatively, as put forward in Ref.~\cite{roman}, one may create artificial topological \lq\lq band-structures\rq\rq\ in multi-terminal Josephson junctions, where the superconducting phases of the reservoirs play the role of quasi-momenta that determine the energy spectrum of the Andreev bound states (ABS) at the junction. In particular, it was shown that four-terminal junctions can be viewed as effectively three-dimensional \lq\lq materials\rq\rq\ that may host zero-energy Weyl singularities. In that case, the three independent superconducting phase differences between the reservoirs yield three control parameters that allow one to realize a zero-energy state without having to fine-tune the scattering properties of the junction.

Here we show that Weyl singularities may also be realized in three-terminal structures when adding a magnetic flux through the junction area as a third control parameter to the two superconducting phases differences. Such a setup might be simpler to realize experimentally  than four-terminal structures, and possibly yield a larger voltage range for observing topological  signatures in transport measurement~\cite{roman,erik}. Furthermore, it displays some interesting features -- concerning which states are responsible for the topological properties -- that distinguish it from the four-terminal case as we will discuss below.

\begin{figure}
\resizebox{.35\textwidth}{!}{\includegraphics{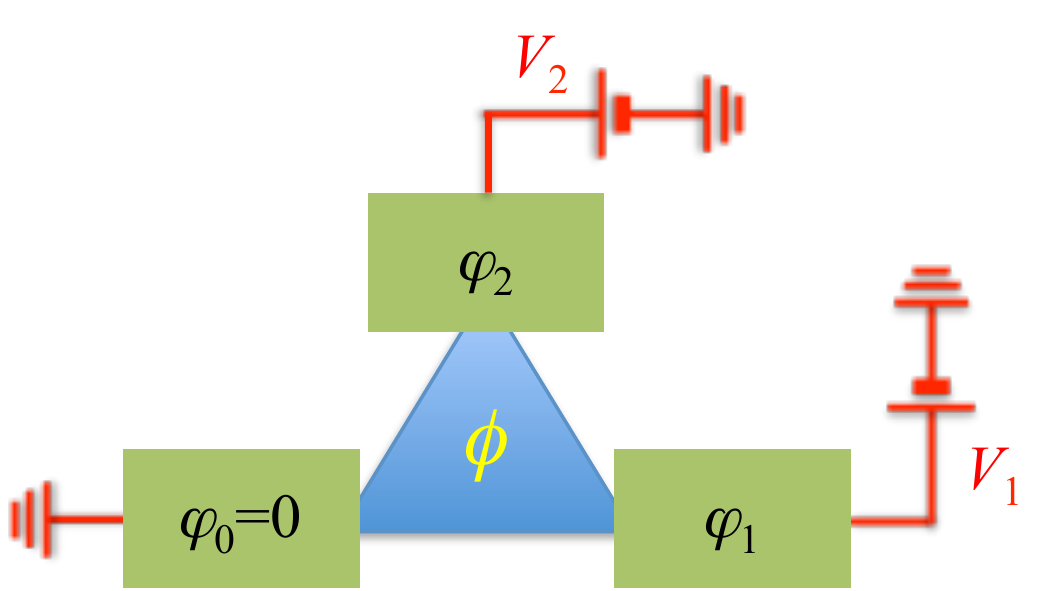}}
\caption{\label{fig-setup} Schema of the setup. Three superconducting reservoirs are connected via a normal scattering region that may be threaded by a magnetic flux. To perform a transconductance measurement, two of the terminals are voltage biased.}
\end{figure}

{\em General setup.} We consider three superconducting reservoirs connected via a central normal scattering region, see Fig.~\ref{fig-setup}. Each superconductor is characterized by an order parameter $\Delta_i=\Delta e^{i\varphi_i}$ with amplitude $\Delta$, chosen to be equal in all reservoirs, and phase $\varphi_i$ ($i=0,1,2$). We choose a gauge where $\varphi_0=0$. The central region can be described using a normal-state scattering matrix $\hat S(\phi)$ whose properties can be tuned by applying a magnetic flux $\phi$. The ABS forming at the junction will, thus, depend on three parameters: the phases $\varphi_1$ and $\varphi_2$ as well as the magnetic flux $\phi$. For simplicity, we concentrate on the case where each reservoir is linked by a single channel and the scattering region is small enough such that the energy-dependence of the scattering matrix can be neglected in a first step. 

Assuming spin-rotation invariance, the ABS spectrum is determined by the eigenproblem~\cite{Beenakker1991}
\begin{equation} 
|\psi\rangle=a^2(E)\hat S(\phi)e^{i\hat\varphi}\hat S^*(\phi)e^{-i\hat\varphi}|\psi\rangle,
\end{equation}
where $a(E)=E/\Delta-i\sqrt{1-(E/\Delta)^2}$ is the Andreev reflection amplitude and $\hat\varphi$ is a diagonal matrix assigning to each reservoir its superconducting phase. Furthermore, $|\psi\rangle$ is a 3-vector of the amplitudes of outgoing wavefunctions at the junction in the electron-like sector, from which the rest of the wavefunction can be reconstructed.

The ABS spectrum can, thus, be found by determining the eigenvalues $\lambda$ of the matrix $\hat A=\hat S(\phi)e^{i\hat\varphi}\hat S^*(\phi)e^{-i\hat\varphi}$, and determining the energy using the condition $a^2\lambda=1$. Due to particle-hole symmetry, one eigenvalue is stuck to the gap edge, $\lambda_0=1$ corresponding to $|E|=\Delta$, whereas the two other eigenvalues $\lambda_\pm=\exp[\pm i\alpha]$ correspond to energies $E_{\rm ABS}=\pm\Delta\cos(\alpha/2)$. Solving the characteristic equation ${\rm Det}[\hat A-\lambda\mathbbm{1}]=0$, one finds
\begin{equation}
E_{\rm ABS}=\pm\frac\Delta2\sqrt{1+{\rm Tr}\big[\hat A\big]}.
\end{equation}
Thus, a Weyl point at $E=0$ appears if ${\rm Tr}[\hat A]=-1$ for some value of $(\varphi_1,\varphi_2,\phi)$. As there are three tuning parameters, such a crossing between two levels may be achieved without fine-tuning the scattering matrix $\hat S(\phi=0)$.  Furthermore, each zero-energy crossing is characterized by a topological charge $q=\pm 1$, and a multiple of four such crossings -- with zero total charge -- are expected in the volume defined by $0\leq \varphi_1,\varphi_2,s<2\pi$.

{\em Scattering matrix from Hamiltonian.} While the dependence on the superconducting phases is explicit in the expression for $\hat A$, we need a model to determine the dependence of the normal-state scattering matrix $\hat S$ on the magnetic flux.

The simplest toy model consists of reservoirs that are connected via a single site with on-site energy $U_i$ and a hopping matrix element $t_{ij}$ between reservoirs:
\begin{eqnarray}
\label{eq:toy}
H&=&H_{\rm reservoirs}
+\sum_{i,\sigma} U_i\psi_{i\sigma}^\dagger(0)\psi_{i\sigma}(0)\\
&&\qquad\qquad\qquad
+\sum_{i<j,\sigma} \left\{t_{ij}\psi_{j\sigma}^\dagger(0)\psi_{i\sigma}(0)+{\rm h.c.}\right\},
\nonumber
\end{eqnarray}
where $\psi_{i\sigma}(0)$  is an annihilation operator of an electron with spin $\sigma$ in lead $i$ at the junction. In the absence of a magnetic flux, the hopping matrix elements can be choosen real such that $t_{ij}=t_{ji}$. Due to a magnetic flux $\phi$ through the junction, the hopping matrix elements acquire a phase $\theta_{ij}$ such that the total phase $\theta_{12}+\theta_{23}+\theta_{31}=\phi/\phi_0\equiv s$, where $\phi_0=h/e$ is the flux quantum. We may choose $\theta_{12}=\theta_{23}=\theta_{31}=s/3$. A different repartition can be compensated by adjusting the superconducting phases. Thus, the coupling term takes the form
\begin{equation}
\hat W=\begin{pmatrix}U_1&t_{12}e^{is/3}&t_{31}e^{-is/3}\\t_{12}e^{-is/3}&U_2&t_{23}e^{is/3}\\t_{31}e^{is/3}&t_{23}e^{-is/3}&U_3\end{pmatrix}.
\end{equation}
In the wide band limit, where the density of states $\nu$ in the reservoirs can be taken as a constant, the scattering matrix is obtained as~\cite{Ziman}
\begin{equation}
\hat S=(1-i\pi\nu\hat  W)^{-1}(1+i\pi\nu \hat W).
\end{equation}

\begin{figure}
\resizebox{.3\textwidth}{!}{\includegraphics{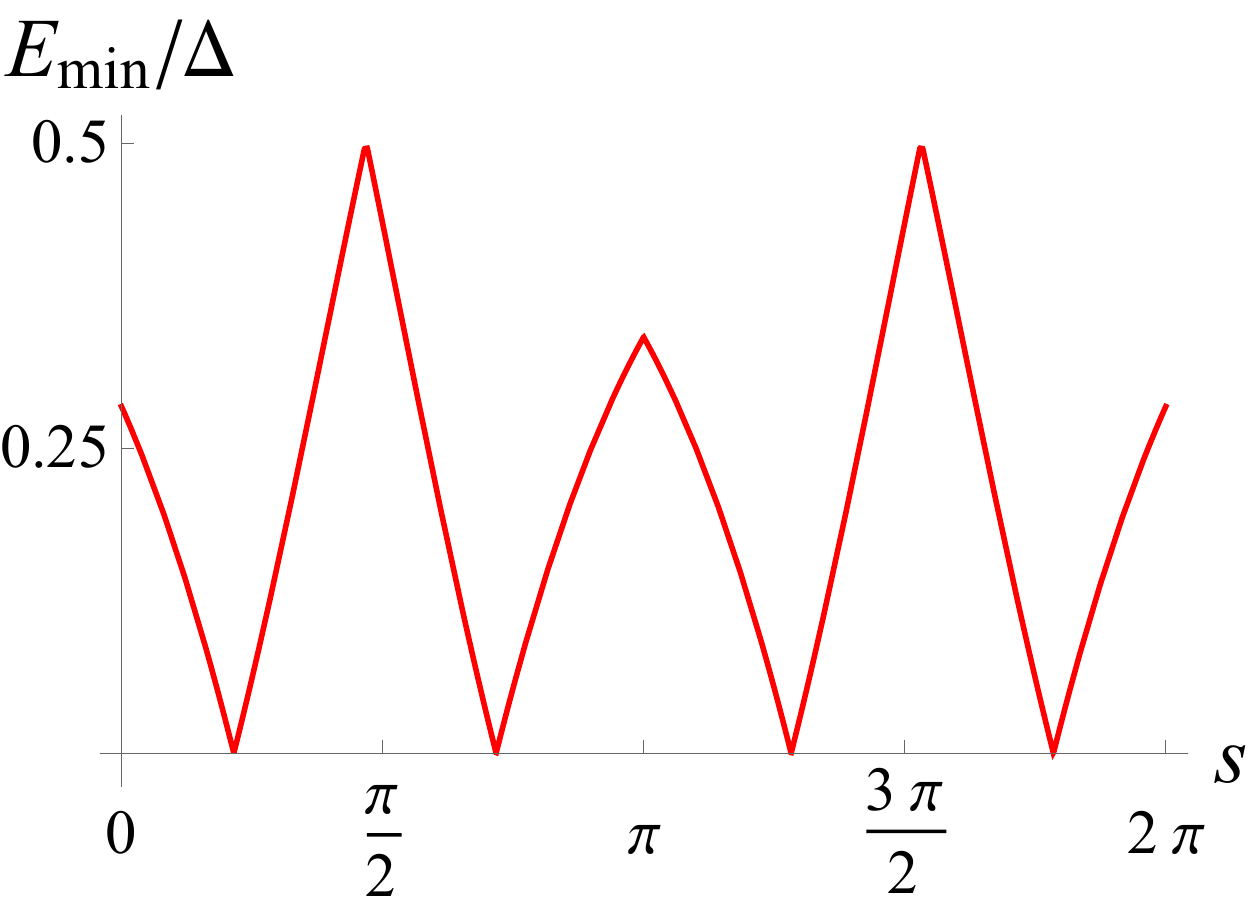}}
\caption{\label{fig-spectrum} Minimal energy $E_{\rm min}(s)=\min_{\varphi_1,\varphi_2} E_{\rm ABS}(\varphi_1,\varphi_2,s)$ for the symmetric case with $U=0.1$ and $t=1$. We find four Weyl points at $s=0.676, 2.254,4.029,5.608$. The energy gap is largest at $s=1.479,4.804$, where $E_{\rm min}/\Delta\approx0.49$.}
\end{figure}

{\em Symmetric case.}  The model simplifies significantly when one chooses a completely symmetric model where all the on-site energies are the same, $\pi\nu U_i\equiv U$, and all the hoppings are the same, $\pi\nu t_{ij}\equiv t$. In that case, the scattering matrix can be diagonalized by the unitary transformation $\hat  S\to \hat S_D=\hat {\cal U}^\dagger \hat S\hat {\cal U}$, where
\begin{equation}
\hat  {\cal U}=\frac1{\sqrt3}\begin{pmatrix}1&1&1\\1&e^{i2\pi/3}&e^{-i2\pi/3}\\1&e^{-i2\pi/3}&e^{i2\pi/3}\end{pmatrix}
\end{equation}
and $\hat S_D=(1-i\hat D)^{-1}(1+i\hat D)$ with
\begin{equation}
\hat D=\begin{pmatrix}U+2t\cos\frac s3&&\\&U+2t\cos\frac{s+2\pi}3&\\&&U+2t\cos\frac{s-2\pi}3\end{pmatrix}.
\end{equation}
In particular, we can show that the minimal energy at a given flux is obtained for $\varphi_1=-\varphi_2=2n\pi/3$ with $n=0,1,2$. For these values of the phases, the eigenvalues of $\hat A$ are given as $\lambda_0=1$, $\lambda_+ = [S_D^{22}(s+2n\pi/3)]^*S_D^{33}(s+2n\pi/3)$, and $\lambda_-=\lambda_+^*$. The energy vanishes when $\lambda_\pm=-1$ at 
\begin{equation}
\cos\frac{s+2n\pi}3=\frac U{4t}\pm\frac1{4t}\sqrt{12t^2-3U^2-4}.
\end{equation}
Thus, Weyl points exist if $12t^2-3U^2-4>0$ or, equivalently, $|t|>\sqrt{(3U^2+4)/12}$. When $|t|=\sqrt{(3U^2+4)/12}$, two Weyl points with opposite charges merge at $s=3\arccos\sqrt{3U^2/(12U^2+16)}$. At $t=(U\pm\sqrt{9U^2+8})/4$, two Weyl points with the same charge coexist in the $s=0$--plane, whereas at $t=(-U\pm\sqrt{9U^2+8})/4$, two Weyl points with the same charge coexist in the $s=\pi$--plane. In the limit $t\to\infty$, the Weyl points approach $s=(2n+1)\pi/2$. The minimal energy as a function of flux for $U=0.1$ and $t=1$ is shown in Fig.~\ref{fig-spectrum}.

{\em Quantized transconductance (numerics).} As shown in Refs.~[\onlinecite{roman,erik}], the presence of Weyl singularities leads to a quantized transconductance when voltage-biasing two of the superconducting reservoirs and sweeping a control parameter. In particular,
\begin{equation}
G_{\alpha\beta}=-\frac{4e^2}h C_{\alpha\beta},
\end{equation}
where the Chern number is given as
\begin{equation}
C_{\alpha\beta}=\frac i{2\pi}\sum_n\int\limits_0^{2\pi}d\varphi_\alpha\int\limits_0^{2\pi}d\varphi_\beta\;\langle\partial_{\varphi_\alpha} \psi_n|\partial_{\varphi_\beta} \psi_n\rangle,
\end{equation}
summing is over all (bound and continuum) states $|\psi_n\rangle$ with negative energy.

We use the same method as described in Ref.~[\onlinecite{erik}] to compute the transconductance in the three-terminal case, using the magnetic flux as the control parameter. Namely we calculate the currents flowing through the setup at arbitrary voltage biases with the Landauer-B{\"u}ttiker scattering formalism extended to superconducting hybrid structures~\cite{averin-bardas}. The quantized transconductance is due to the non-trivial Chern number of the states below the Fermi level and requires averaging over the entire space of phases $(\varphi_1,\varphi_2)$. This may be achieved by applying incommensurate voltages $V_1$ and $V_2$. For computational purposes, however, this is not ideal. Alternatively, the computation may be done either by using two commensurate voltages $V_1=n_1V$ and $V_2=n_2V$, and then averaging of the phase offset between the phases $\varphi_1$ and $\varphi_2$, or by using a single voltage $V_1=V$ and averaging over the phase $\varphi_2$. While in Ref.~[\onlinecite{erik}] the first option was chosen, here we show results using the second option, which requires less computing time. In order to account for inelastic relaxation, we introduce a Dynes parameter~\cite{Dynes}.

\begin{figure}
(a)\!\!\!\!\!\!\!\!\!\!\!\resizebox{.5\textwidth}{!}{\includegraphics{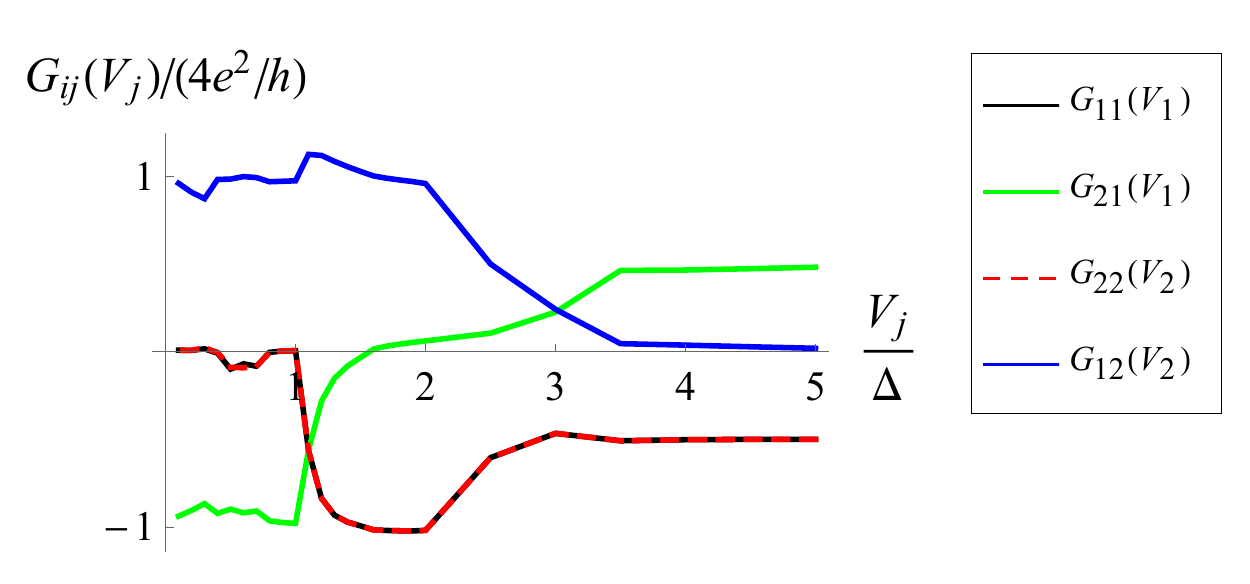}}
\\
(b)\!\!\!\!\!\!\!\!\!\!\!\resizebox{.5\textwidth}{!}{\includegraphics{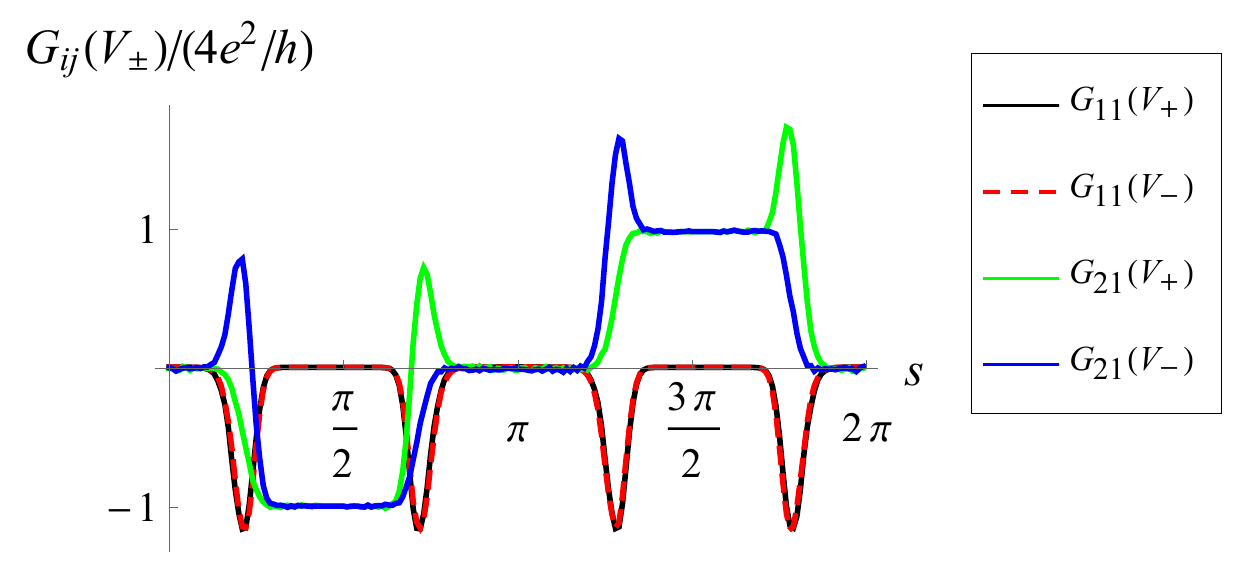}}
\caption{\label{fig-numerics} Conductances for the symmetric case with $U=0.1$ and $t=1$. Here the Dynes parameter is $\Gamma/\Delta=0.01$. (a) Conductances $G_{11}(V_1)$ and $G_{21}(V_1)$ as well as  $G_{22}(V_2)$ and $G_{12}(V_2)$ as a function of the voltage at $s=\pi/2$. (b) Conductances $G_{11}$ and $G_{21}$ as a function of the magnetic flux at $eV_\pm/\Delta=\pm 0.02$. While the plateau values are the same for both signs of the voltage, the dissipative regime close to the Weyl points strongly depends on how the bias is applied.}
\end{figure} 

In Fig.~\ref{fig-numerics}, the conductances, $G_{ij}=\partial I_i/\partial V_j$, are shown, both as a function of voltage at fixed flux [Fig.~\ref{fig-numerics}(a)] and as a function of flux at fixed voltage [Fig.~\ref{fig-numerics}(b)]. The quantized transconductance plateaus at low voltage are clearly visible. Close to the Weyl points, the dissipation is large and the value of the transconductance strongly depends on how the voltages are applied, as can be seen by comparing the results for $V$ and $-V$ in Fig.~\ref{fig-numerics}(b).

\begin{figure}
(a)\resizebox{.2\textwidth}{!}{\includegraphics{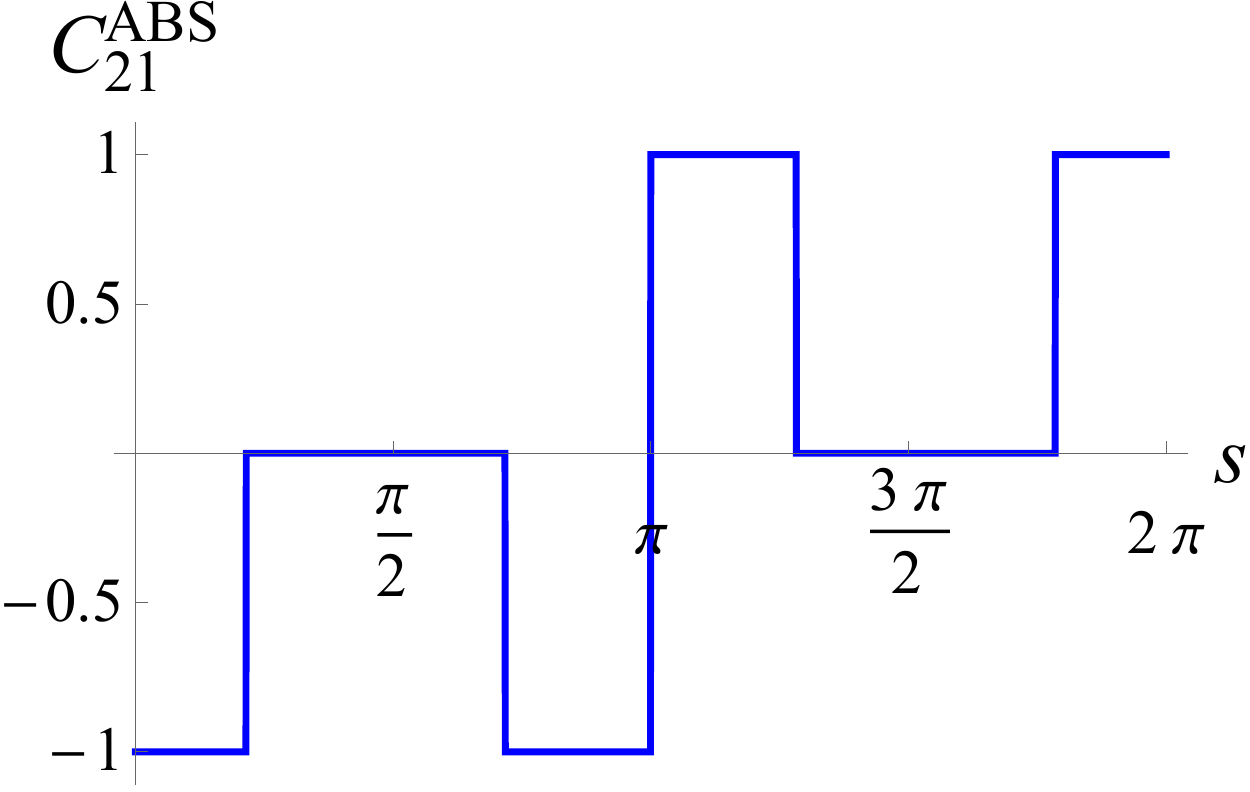}}\hfill(b)\resizebox{.225\textwidth}{!}{\includegraphics{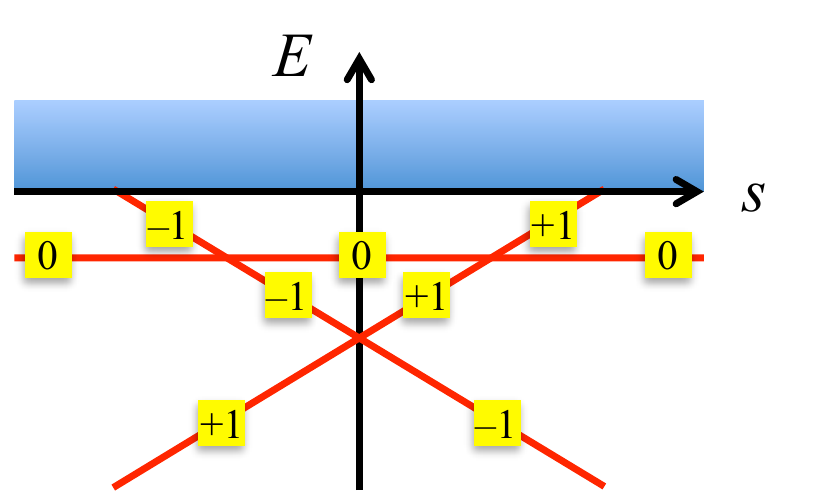}}
\caption{\label{fig-chern} (a) Chern number $C_{21}^{\rm ABS}$ of the ABS with negative energy for the symmetric case with $U=0.1$ and $t=1$. (b) Flux dependence of the ABS at $\varphi_1=\varphi_2=0$  close to $\Delta$ when a weak energy dependence of the normal-state scattering matrix is taken into account. The charge-2 Weyl point at $s=0$ leads to a non-trivial Chern number of the lowest-energy state at $s\neq0$. }
\end{figure}

{\em Chern number of the ABS.} Interestingly the quantized transconductance is not directly linked with the Chern number of the ABS. Figure~\ref{fig-chern}(a) shows the Chern number of the negative energy state: it is zero in the range of $s$ where the transconductance is finite whereas it is non-trivial in the range of $s$ where the transconductance vanishes. Its jumps $\delta C_{21}^{\rm ABS}=\pm1$ occur in planes containing zero-energy Weyl crossings with topological charges $q=\pm 1$. Furthermore, at flux $s=0,\pi$, the Chern number displays jumps of $\delta C_{21}^{\rm ABS}=\pm2$. Such jumps are only possible when closing a gap in the spectrum. This is indeed the case at these values of the flux, where the ABS reaches the continuum at $\pm\Delta$. We thus conclude that the quantized transconductance is due to the properties of the continuum states.

{\em Energy-dependent scattering matrix.} To further explore what happens at flux $s=0$, we take into account a weak energy dependence of the scattering matrix $\hat S$ which allows one to separate the ABS from the continuum at phases $\varphi_1=\varphi_2=0$. As the energy dependence of the scattering matrix is on the scale of the Thouless energy $E_{\rm Th}$, we approximate $\hat S\approx \hat S_0+\Delta\partial_E \hat S$, where the second term is a small correction for $\Delta\ll E_{\rm Th}$ corresponding to a sufficiently short normal region.
Expanding the matrix $\hat A$ around $\varphi_1=\varphi_2=s=0$~\cite{tomohiro}, we find
\begin{eqnarray}
\delta \hat A&=&\Delta (\partial_E \hat S\hat S_0^\dagger+\hat S_0\partial_E \hat S^*)\\
&&+i\hat S_0[\hat \varphi ,\hat S_0^\dagger]+s(\partial_s\hat S\hat S_0^\dagger+\hat S_0\partial_s \hat S^*),\nonumber
\end{eqnarray}
where we took into account that $\hat S_0(\phi=0)=\hat S_0^T(\phi=0)$. Using $a^2(E_{\rm ABS})\approx1-2i\sqrt{2\varepsilon}$ with $\varepsilon=1-E_{\rm ABS}/\Delta$, the eigenvalues close to $\Delta$ are obtained by solving
$\delta \hat A|\psi\rangle = 2i\sqrt{2\varepsilon}|\psi\rangle$.
The equation may be rewritten as $\hat {\cal H}|\tilde\psi\rangle = \sqrt{2\delta\epsilon}|\tilde\psi\rangle$, where the effective Hamiltonian $\hat {\cal H}$ is given as
\begin{equation}
\label{eq:Q}
\hat{\cal H}=\hat S_0^{1/2}\left\{\Delta \hat Q+\frac12[\hat \varphi ,\hat S_0^\dagger]\hat S_0-is\hat S_0^\dagger\partial_s\hat S\right\}\hat S_0^{\dagger\,1/2}.
\end{equation}
Here $\hat Q=-i\hat S_0^\dagger\partial_E \hat S$ is the Wigner-Smith time delay matrix~\cite{wigner,smith}. It is Hermitian and positive-definite, due to the causality of the scattering matrix. 

The advantage of the formulation \eqref{eq:Q} is that the time-reversal invariant and non-invariant contributions to $\hat{\cal H}$ are real and pure imaginary, respectively. Diagonalizing the time-reversal invariant part with an orthogonal matrix, we can rewrite the effective Hamiltonian as
\begin{equation}
\label{eq:Heff}
\hat {\cal H}_D=\left(\begin{array}{ccc}
\Delta \tau_1 & -i \chi_1 & -i \chi_2\\
i \chi_1 & \Delta\tau_2 & -i \chi_3 \\
i \chi_2 &i \chi_3& \Delta\tau_3
\end{array}\right).
\end{equation}
The eigenvalues $\tau_{1,2,3}>0$ of $\hat S_0^{1/2}\hat Q \hat S_0^{\dagger\,1/2} $, thus, result in the detachment of three ABS from the continuum with energies $\Delta[1-(\Delta\tau_i)^2/2]$. In general, the three eigenvalues are all different, though in the symmetric case considered above, we find two degenerate eigenvalues,
$\tau_1=2\partial_ED_{11}(s\!=\!0)/d_1$ and $\tau_{2/3}=2\partial_ED_{22}(s\!=\!0)/d_2$,where $d_i=1+D_{ii}^2(s\!=\!0)$. In that case, while the non-degenerate eigenvalue does not vary with flux, the other two split linearly as $\pm \chi_3$ with $\chi_3=2st/(\sqrt{3}d_2)$. By contrast, the dependence on the phases due to $\chi_{1/2}=[(\sqrt{3}+1)\varphi_{1/2} -(\sqrt{3}-1)\varphi_{2/1}]t/(2\sqrt{d_1d_2})$  is quadratic. 

The linear dependence on $s$ combined with the quadratic dependence on phases indicates~\cite{c2-weyl} that the two levels that cross at $\varphi_1=\varphi_2=s=0$ form a Weyl point with topological charge $\pm 2$. Two additional Weyl crossings with topological charges $\mp 1$ occur at $\varphi_1=\varphi_2=0$ and $s=\pm \sqrt{3}\Delta(\tau_1-\tau_2)d_2/(2t)$. Thus, one can assign Chern numbers as shown in Fig.~\ref {fig-chern}(b) to the different levels (assuming $\tau_1<\tau_2$ for concreteness). While the levels closer to the gap will merge with the continuum when further increasing the phases or the flux, the Chern number of the lowest level is the one we observe for the ABS that exists at all values of $(\varphi_1,\varphi_2)$. In the non-symmetric case, the Weyl point at $s=0$ with charge $\pm2$ splits into two Weyl points of equal charge $\pm1$ at $s=\pm s_0$~\cite{footnote}. 
The Chern numbers at larger values of $s$ remain the same.

Since the total Chern number carried by the occupied states remains constant, no jump in $G_{12}$ is expected because of the crossings of ABS away from $E=0$.

{\em Junction through a chaotic dot.} 
Similar results to those of the toy model \eqref{eq:toy} are found for a junction through a chaotic dot threaded by a flux $\phi$. In the spirit of random matrix theory~\cite{RMT}, we can generate the scattering matrix for such a junction using the formula~\cite{brouwer}
\begin{equation}
\label{eq:S-RMT}
\hat S(x)=\hat{\Sigma}_{11}+\hat{\Sigma}_{12}[\mathbbm{1}-\hat r(x)\hat{\Sigma}_{22}]^{-1}
\hat r(x)\hat{\Sigma}_{21}.
\end{equation}
Here $\hat \Sigma_{ij}$ are the four blocks of a $(3+N)\times (3+N)$ matrix with $N\gg 1$ drawn out of the circular orthogonal ensemble, {$\hat r(x)=\exp(x\hat B)$, where $\hat B$ is an arbitrary real and antisymmetric $N\times N$ matrix with ${\rm Tr}[ \hat B^2]=1$, and $x$ is a real parameter such that the distribution of $\hat S(x)$ crosses over from the circular orthogonal ensemble at $x=0$ to the circular unitary ensemble at $x\gg 1$. Furthermore, the relation between $x$ and $\phi$ is given by $x\sim s\sqrt{E_{\rm Th}/\delta}$,} where $\delta$ is the mean level spacing in the dot. In contrast to the toy model, the scattering matrix \eqref{eq:S-RMT} is not periodic in flux.

\begin{figure}
\resizebox{.4\textwidth}{!}{\includegraphics{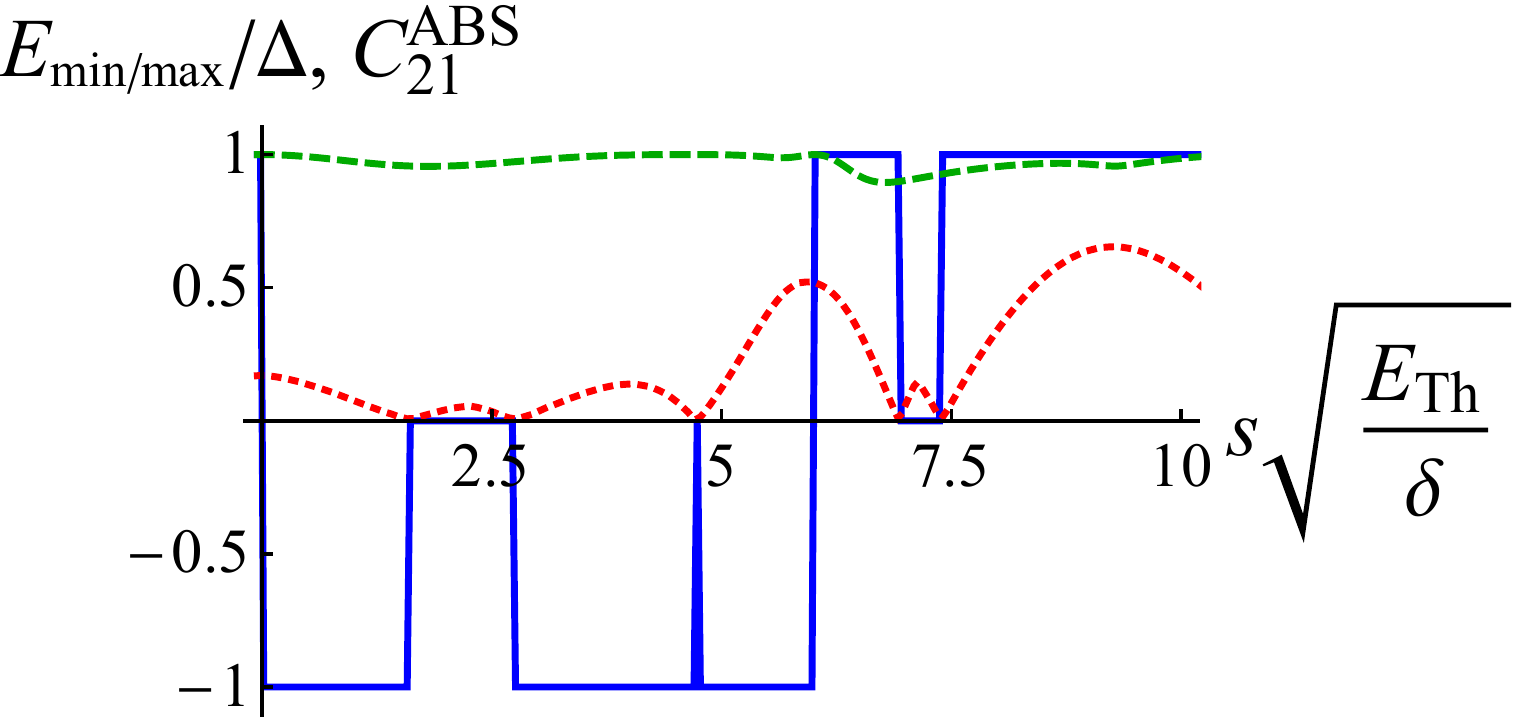}}
\caption{\label{fig-chern-RMT} Chern number $C_{21}^{\rm ABS}$ of the ABS with negative energy (solid line), Andreev gap $E_{\rm min}$ (dotted line), and maximal ABS energy $E_{\rm max}$ (dashed line) as a function of {
$x\sim s\sqrt{E_{\rm Th}/\delta}$} for a three-terminal junction through a chaotic dot described by a scattering matrix generated randomly according to Eq.~\eqref{eq:S-RMT}.
}
\end{figure} 

The condition for a single ABS to detach from the continuum within that model, $\Delta\lesssim\delta\ll E_{\rm Th}$, translates into an estimate for the size of the dot: $L\lesssim(\xi^2\lambda_F)^{1/3}$, where $\lambda_F$ is the Fermi wavelength and $\xi$ is the superconducting coherence length, which is possible to realize with metallic dots. We found that most scattering matrices (with $N=20$) generated this way yield zero-energy crossings for values of $x$ of order 1. We show the Chern number and minimal ABS energy as a function of $x$ for one of them in Fig.~\ref{fig-chern-RMT}. 
 As for the toy model, we observe jumps $\delta C_{21}^{\rm ABS}=\pm 1$ when the Andreev gap closes, as well as jumps $\delta C_{21}^{\rm ABS}=\pm 2$, which can be related with the ABS touching the continuum.
 
  Before concluding, let us note that we neglected the effect of the magnetic field on the superconducting reservoirs. This is possible only if the magnetic field is confined to the junction area, or if the magnetic field is smaller than the upper critical field $H_{c2}\sim \phi_0/\xi^2$, thus limiting the flux range.

{\em Conclusion.} We have shown that Weyl singularities appear in the ABS spectrum of three-terminal Josephson junctions when using a magnetic flux through the junction as a tuning parameter. The presence of these Weyl singularities entails a quantized transconductance when two of the terminals are voltage-biased, that displays jumps as a function of the magnetic flux. Interestingly this quantized transconductance cannot be ascribed to the ABS alone, but involves the continuum states. On the other hand, the ABS may carry a non-trivial Chern number even though the junction behaves trivially.

\acknowledgments
We appreciate fruitful discussions with Yuli Nazarov. This work was supported by the Nanosciences Foundation in Grenoble (France).


\begin{thebibliography}{99}
\bibitem{footnote}
In lowest order in perturbation theory, the projection of Eq.~\eqref{eq:Heff} on the subspace of degenerate states of the symmetric system at $\chi_1=\chi_2=\chi_3=0$ yields an effective $2\times 2$ Hamiltonian 
$${\cal H}_{\rm eff}=\frac\Delta2{(\tilde \tau_2+\tilde \tau_3)}+\chi_1\chi_2\sigma_x+\chi_3\sigma_y+\frac\Delta2{(\tilde \tau_2-\tilde \tau_3)}\sigma_z,$$ 
where $\tilde \tau_{2/3}=\tau_{2/3}+\chi_{2/3}^2/\Delta$ and $\sigma_{x,y,z}$ are Pauli matrices. It yields two Weyl points with the same charge at $\chi_{1/2}=\pm\sqrt{\Delta|\tau_2-\tau_3|}$ and $\chi_{2/1}=s=0$ if $\tau_2\lessgtr\tau_3$. Their shift away from $s=0$ occurs in higher order in the asymmetry.
\end{thebibliography}
\end{document}